\documentclass[final,authoryear,3p,times,twocolumn]{elsarticle}
\usepackage{times,amsmath,amsfonts,amssymb,graphicx,bm,fancyhdr,hyperref}
\def\refname{References}

\newcommand{\DD}[2]{\frac{\partial #1}{\partial #2}}
\newcommand{\DDD}[2]{\frac{\partial^2 #1}{\partial #2^{2}}}

\DeclareMathOperator{\tr}{tr}

\DeclareMathOperator{\diverg}{div}

\DeclareMathOperator{\grad}{grad}

\DeclareMathOperator{\dd}{d\!}
\begin{document}

\begin{frontmatter}
\title{Frame Indifferent Formulation of Maxwell's Elastic-Fluid Model and the Rational Continuum Mechanics of the Electromagnetic Field} 

\author[cic]{C. I. Christov}\ead {christov@louisiana.edu}
\address[cic]{Dept. of Mathematics, University of Louisiana at lafayette, P.O. Box 41010, Lafayette, LA 70504-1010}
\begin{abstract}
We show that the linearized equations of the incompressible elastic  medium admit a `Maxwell form'  in which the shear component of the stress vector plays the role of the electric field, and the vorticity plays the role of the magnetic field. Conversely, the set of dynamic Maxwell equations are strict mathematical corollaries from the governing equations of the incompressible elastic medium. This suggests that the nature of `electromagnetic field' may actually be related to an elastic continuous medium. The analogy is complete if the medium is assumed to behave as  fluid  in shear motions, while it may still behave as elastic solid under compressional motions. Then the governing equations of the elastic fluid are re-derived in the Eulerian frame by replacing the partial time derivatives by the properly invariant (frame indifferent) time rates. The `Maxwell from' of the frame indifferent formulation gives the frame indifferent system that is to replace the Maxwell system. This new system includes terms already present in the classical Maxwell equations,  alongside terms that are the progenitors of the Biot--Savart, Oersted--Ampere's, and Lorentz--force laws. Thus, a frame indifferent (truly covariant) formulation of electromagnetism is achieved from a single postulate: the electromagnetic field is a kind of elastic (partly liquid, partly solid) continuum. 
\end{abstract}
\end{frontmatter}

%\maketitle
\hyphenation{e-lec-tro-mag-net-ism}

%\setlength{\bibspacing}{\baselineskip}

%\tableofcontents
%\newpage

%\bibliographystyle{unsrt}

%\setcounter{page}{3}

%\begin{center}
%{\LARGE\sc Continuum-Mechanics Approach to Unified Field Theory}
%\end{center}

\section{Introduction}

Classically, wave motion in elastic media is a phenomenon associated with either/both the transverse or longitudinal vibrations; if there is a wave, then something material should be waving. This notion led 19th century scientists to introduce the concept of the luminiferous medium (field, aether, etc.).  The first attempt to explain the propagation of light as a field phenomenon was by Cauchy circa 1827 (see the account in \citep{Whittaker1}), who postulated the existence of an elastic continuum through which light propagates as a shear wave.  Subsequently came the contributions of Faraday and Ampere, which eventually led to the formulation of the electromagnetic model as known today. The crucial advance in understanding the phenomena of   electromagnetism phenomena was achieved, however, when \citet{JCM2} added the  `displacement current', $\frac{\partial \bm{E}}{\partial t}$, to Ampere's law. This term was similar to the time-derivative of the stress in his constitutive relation for elastic gases \citep{MaxwellGases}.  Since the electric field vector is a clear analogue of the stress vector in continuum mechanics, one can say that Maxwell postulated an elastic constitutive relation by adding the displacement current to Ampere's law. Indeed, the new term transformed the system of equations (already established at that time in electrostatics) into a hyperbolic system with a characteristic speed of shear wave  propagation.  Maxwell and Hertz identified the characteristic speed with the speed of light.

However, soon after Maxwell formulated his equations, it was discovered that his model was not invariant with respect to translational motion of the coordinate frame.   \cite{Hertz} realized that the cause of non-invariance was the use of partial time derivatives. He proposed to use the convective time derivative \emph{in lieu} of the former.  The Maxwell--Hertz equations (called also `progressive wave equations') are clearly the  correct model for electromagnetic phenomena in moving bodies.  But the primordial question is whether the progressive-wave equations can be construed to hold also \emph{in vacuo}. The answer is obviously in the affirmative if one accepts that what is currently called `physical vacuum' must be regarded as a material  continuum.  

Voigt (see \cite{ErnstHsu}), \citet{Larmor}, and independently Lorentz, spotted the fact that the wave equation can be made invariant if, in the moving frame, the time variable is changed together with the spatial variables. Nowadays, this is known as the Lorentz transformation. The success of the latter stems from the fact that it tacitly restores some parts of the convective derivative, i.e. it emulates the material invariance for  non-deformable frames in rectilinear motion \citep{Chri_FP,Lumini_WCNA}. Although researchers usually speak about Lorentz covariance as a general covariance, it has to be pointed out that the Lorentz transformation has no meaning for accelerating frames (nor for generally deforming frames). Hence the search for the truly covariant formulation should continue. 

The present paper summarizes the efforts from the last decade and a half towards identifying the mechanical construct behind the phenomenon of what is termed `electromagnetic field'. We show here that the Maxwell equations and the laws of electromagnetism (Biot--Savart, Oersted--Ampere, and Lorentz--force laws), all can be derived from the governing equations of Maxwell's elastic fluid, that includes relaxation of the stress. In doing so, we also present a concise frame indifferent formulation of the Maxwell elastic fluid model.

\setcounter{equation}{0}
\section{Maxwell Form of the  Equations of Incompressible Elastic Medium}\label{sec:Maxwell}

The linearized governing equations of elastic medium are valid only for infinitesimal deformations, when the referential description $\bm{X}$ and spatial configuration $\bm{x}$  coincide. The linear constitutive relationship for an elastic body relates the stress tensor $\hat \sigma$ (we reserve the notation $\sigma$ for the deviatoric stress tensor) to the deformation tensor $e$ via the generalized Hooke's law
\begin{equation}
\hat \sigma  = \lambda \tr(e)+ 2\eta e, \quad  e:= \tfrac{1}{2}[\nabla \bm{u} +(\nabla \bm{u})^\intercal],
\end{equation}
where $\tr(e)$ is the trace of the deformation tensor, and the superscript $\intercal$ denotes the transpose. The above constitutive law yields the so-called Navier equations (see, e.g., \citep[pg.~117]{Segel} for the displacement vector $\bm{u}(\bm{x};t)$:
\vspace{-0.1in}\begin{multline}
\mu\DDD{\bm{u}}{t} = (\lambda +\eta)\nabla (\nabla \cdot \bm{u}) + \eta \nabla^2 \bm{u}\\
=(\lambda +2\eta)\nabla (\nabla \cdot \bm{u})- \eta\nabla\times\nabla \times\bm{u},
\label{eq:linearized}
\end{multline}
where $\lambda$ and $\eta$ are the Lam\'e coefficients. Here one can use the `nabla' operator $\nabla$ because the equations are written in the current configuration.

The speeds of propagation of shear and compressional disturbances, are given respectively by
\begin{equation}
c^2 =  \frac{\eta}{\mu} ,
\ \ c^2_s = \frac{2\eta + \lambda}{\mu}, \ \ \delta = \frac{\eta}{2\eta + \lambda} =\frac{ c^2}{c_s^{2}},
\label{eq:speeds}
\end{equation}
where the ratio $\delta$ is introduced for convenience.

In a compressible elastic medium, both the shear and the dilational/compressional waves should be observable. Since the groundbreaking works of Young and Fresnel, it is well established that electromagnetic waves (e.g., light) are a purely transverse (shear) phenomenon. This observation requires us to reduce the complexity of the model and to find a way to eliminate the dilational modulus $\lambda$. Cauchy assumed that $\lambda=0$ and ended up with the theory of so-called `volatile aether' (see \cite{Whittaker1}). Upon a closer examination, we found that such an approach cannot explain Maxwell's equations.

Let us now assume that the dilational waves are not observable because the other extreme situation is at hand: $\lambda \gg \eta$ which is equivalent to  $\delta \ll 1$ or $\delta^{-1} \gg 1 $. It is convenient to rewrite Eq.~\eqref{eq:linearized} in terms of the speed of light, $c$, and parameter $\delta$, namely
\begin{equation}
\delta\Big(c^{-2}\DDD{\bm{u}}{t} + \nabla\times\nabla \times\bm{u}\Big)
=\nabla (\nabla\cdot\bm{u}),
\label{eq:linearized1}
\end{equation}
and to expand the speed of light $c$,  displacement $\bm{u}$, and velocity $\bm{v}:=\partial\bm{u}/\partial t$  into asymptotic power series with respect to $\delta$, namely
\vspace{-0.15in} \begin{align}
c &= c_0 + \delta c_1 + \cdots \ , \nonumber\\
\bm{u} &= \bm{u}_0 + \delta \bm{u}_1 + \cdots \ ,\label{eq:21}\\
\bm{v} &= \bm{v}_0 + \delta \bm{v}_1 + \cdots \ . \nonumber
\end{align}
Introducing \eqref{eq:21} into \eqref{eq:linearized1} and combining the terms with like powers we obtain for the first two terms
\begin{subequations}
\begin{align}
\nabla(\nabla \bm{\cdot} \bm{u}_0) & = 0 \>, \label{eq:22} \\
c_0^{-2} \DDD{\bm{u}_0}{t}
+ \nabla \times \nabla \times \bm{u}_0 &=
\nabla (\nabla \bm{\cdot} \bm{u}_1) \stackrel{\mathrm{def}}{=} - \frac{1}{\mu c_0^2} \nabla p\>,
\label{eq:222}
\end{align}
\end{subequations}
where $p$ is introduced for convenience and is the coefficient of the spherical part of the internal stresses. The variable $p$ has dimension of $\mu c^2$ and plays the same role as the pressure in an incompressible medium: $p$  is an implicit function in Eq.~\eqref{eq:222} that provides the necessary degree of freedom  to enforce the satisfaction of the `incompressibility' condition, Eq.\eqref{eq:22}. The latter can also be rewritten as
\begin{equation}
 \nabla \bm{\cdot} \bm{u}_0 = const, \quad \Rightarrow \quad \nabla\cdot \bm{v}_0=0,
\label{eq:solen}
\end{equation}
which requires that the velocity field be solenoidal within the zeroth-order of approximation of the small parameter $\delta$. From now on, the subscript `$0$' will be omitted form the variables without fear of confusion.

Now, Eq.~\eqref{eq:222} can be rewritten as
\vspace{-0.05in}\begin{equation}
\mu \DD{\bm{v}}{t} = -\nabla p + \bm{s}, \quad \bm{s} \stackrel{\mathrm{def}}{=} \nabla\cdot \sigma=  - \eta\nabla \times \nabla \times \bm{u},\\
\label{eq:momentum}
\end{equation}
where $\bm{s}$ is the tangential part of the stress vector in the elastic continuum. The normal part of the stress vector is given by the pressure gradient $\nabla p$.

Let us now introduce the following notations
\begin{subequations}\label{eq:elmag_definitions}
\begin{gather}\bm{E} \stackrel{\rm def}{=} -\bm{s}
%= -\nabla\bm{\cdot} \sigma
=  \eta\,\nabla\times ( \nabla \times \bm{u}),\label{eq:E-def}
\\ 
\bm{B} \stackrel{\rm def}{=} \mu\, \nabla \times \bm{v} = \eta\, c^2 \nabla \times \bm{v}, \label{eq:H-def}
\end{gather}
\end{subequations}
and call them the `electric field' and `magnetic induction' vectors. Thus the electric field is the negative tangential stress vector, while magnetic field is the `dynamic' vorticity. Naturally, $\bm{H} \!:=\! \nabla \times \bm{v}$ is the `kinematic' vorticity. In the virtue of the above definition, one has the following equation for the magnetic field
\begin{equation}
\nabla \cdot \bm{B}= 0.
\label{eq:B_diverge}
\end{equation}

Now taking the $curl$ of Eq.~\eqref{eq:momentum} and acknowledging the definitions given in Eqs.~\eqref{eq:elmag_definitions}, we arrive at Faraday's law
\begin{equation}
\nabla \times \bm{E} =  -\DD{\bm{B}}{t}.
\label{eq:faraday}
\end{equation}
On the other hand, taking the time derivative of Eq.~\eqref{eq:E-def}, we obtain  the second of the dynamic Maxwell's equations
\begin{equation}
\DD{\bm{E}}{t}  =  \eta\, \nabla \times (\nabla \times
\bm{v}) \equiv c^2 \nabla \times \bm{B}.
\label{eq:rheology_law}
\end{equation}

The fact that the governing equations of any elastic medium in the linear limit admit a Maxwell form also can be considered as an indication that the electromagnetic field is in itself an elastic medium.   In what follows, we shall call the mechanical object equivalent mathematically to the electromagnetic field the \emph{metacontinuum}, to distinguish  it from continuous media in  technical applications, such as fluids and elastic solids. Note that the inverse of the elastic shear coefficient plays the role of the electric permittivity \emph{in vacuo}, while the density of the metacontinuum acts as the  `magnetic permeability.'

The results of this section unequivocally show that the `field' described by Maxwell's equations is equivalent to an elastic material. To best of author's knowledge, the connection of Maxwell's equation to the equations governing elastic media was first established in \citep{Chri_WS,Chri_CMDS8,Chri_annuary}.\footnote{Similar derivations were proposed, presumably independently, in \citep{Dmitriev}.} The common theme of these earlier papers is that the metacontinuum is considered as an elastic \emph{solid}.  In such a model, no \emph{stationary} magnetic fields can  exist, since no steady velocities are possible for a solid continuum without discontinuities.

\section{The Electromagnetic Field: Liquid or Solid?}

In this section we outline the next decisive step in developing the model: we consider an elastic liquid \emph{in lieu} of an elastic solid. This means that for shear deformations, the metacontinuum must be an elastic \emph{fluid} for which the time derivative of the deviatoric  stress tensor is related to the deviatoric rate of deformation tensor $\chi$ via the relation
\begin{equation}
\frac{\partial \sigma}{\partial t} = \eta \chi, \quad  \chi=   \tfrac{1}{2}(\nabla \bm{v} +\nabla \bm{v}^T) -\nabla\cdot \bm{v}. \label{eq:tensor_rheology}
\end{equation}
This  constituitve relationship (rheology) can be rewritten for the negative stress vector $\bm{E}=- \nabla\cdot \sigma$ and deformation vector, $\bm{d} = \nabla \cdot \chi~ (=- \nabla \times \nabla\times \bm{v}$), since both of these vectors are  the divergences of the respective tensors involved in the elastic rheology, Eq.~\eqref{eq:tensor_rheology}. Then
\begin{equation}
 \tau \frac{\partial \bm{E}}{\partial t} = \zeta\, \nabla \times \nabla \times\bm{v}, 
\ \ \text{or} \ \
\frac{\partial \bm{E}}{\partial t}  = \eta\, \nabla \times \nabla \times\bm{v},
\end{equation}
where $\zeta$ is called `elastic viscosity' by \cite{Joseph},  $\tau$ is the relaxation time of the stresses, and the apparent elastic shear modulus is given by $\eta = \zeta\tau^{-1}$. 
Note that the above elastic-liquid rheology concerns just the shear deformations.  For compressional/dilational motions, the metacontinuum can still behave as a virtually incompressible solid. A more general formulation of the shear part of the constitutive relation would be as in the viscoelastic liquid
%\begin{subequations}\label{eq:rheology}
\begin{equation}
\tau \frac{\partial \bm{E}}{\partial t} + \kappa\tau \bm{E} = \zeta\, \nabla \times \nabla \times \bm{v}, \ \ \text{or} \ \
\frac{\partial \bm{E}}{\partial t}+ \kappa \bm{E} = \eta\, \nabla \times \nabla \times \bm{v},
\label{eq:constitutive_solid_liquid}
\end{equation}
where $\kappa$ can be called the `conductivity' of the viscoelastic liquid. Note that in \citep{Joseph}, the conductivity is set to unity, which precludes treating purely elastic (non-viscous) liquids. Here, we prefer to keep the flexibility offered by the presence of the coefficient $\kappa$. Setting the appropriate terminology is an uneasy task because, when $\kappa\ne 0$, then $\zeta$ does indeed have a meaning of a viscosity coefficient, while for $\kappa=0$, it loses its independent meaning and enters the picture through the  coefficient of apparent shear elasticity $\eta$. 

The case $\kappa\!\ne\! 0$ gives a viscoelastic rheology, but it does \emph{not} lead to a model governed by the Navier-Stokes equations\footnote{Usually referred to as Newtonian liquid.}  with additional elasticity, because no retardation term (time derivative of the deformation tensor/vector) is present. In this sense, adding the conductivity does not introduce dispersive dissipation, but rather a linear attenuation parameterized by the conductivity coefficient. For the effects connected with the attenuation/conductivity we refer the reader to \citep{Chri_ML} and \citep{Harmuth}. The constitutive relationship given in Eq.~\eqref{eq:constitutive_solid_liquid} can be interpreted as  Ohm's law for \emph{vacuo}. Although, this stipulation is made in mainstream texts (e.g., \citep{Joos}), the cause of Ohm's law in matter is not necessarily the intrinsic resistance of the metacontinuum: it is the result of  the thermal fluctuations of the atoms that obstruct the free passage of charges through a conductor. Clearly, a more in-depth argument is needed to justify having Ohm's law in \emph{vacuo}, which goes beyond the scope of the present paper.

It should be pointed out here that Eq.~\eqref{eq:constitutive_solid_liquid} is concerned merely with the rheology for the shear motions. At this point it is not of importance if for dilational / compressional motions the metacontinuum is solid or liquid, provided that the dilational coefficient is much larger than the shear coefficient. Then,
\begin{equation}
\tr(\hat \sigma) = \begin{cases}
(\nu + 2 \zeta) \nabla\cdot \bm{v} & \text{liquid}, \\
(\lambda + 2 \eta) \nabla\cdot \bm{u} & \text{solid}.
\end{cases} \label{eq:dilational}
\end{equation}
%\end{subequations}
The intuitive argument is that if the metacontinuum is a liquid with respect to the dilational motions, it may lose its integrity during the motion. Since there is no information on the electromagnetic field being `ruptured,' we are guided by the above argument and assume that the solid rheology applies to the dilational/compressional motions. Note that in the above notations, $\nu$ is the dilational viscosity coefficient, and $\zeta$ is the shear viscosity coefficient defined in Eq.~\eqref{eq:constitutive_solid_liquid}$_1$.

The system \eqref{eq:constitutive_solid_liquid}, \eqref{eq:dilational}$_2$ specify the combined  constitutive relation for the metacontinuum (electromagnetic field). The closest analogy to an ubiquitous continuous medium is that of jelly or pine pitch. %It is possible to control the properties of the viscoelastic liquid if the agar-gelatin based phantom is used in the experiments (see, e.g., \citep{Catheline}). 
If compression/dilation waves are sent through the metacontinuum, it behaves as a elastic solid with very large dilational modulus, while if a shear deformation is applied, it flows as a incompressible liquid.

\section{Euler Variables: the Frame Indifferent Model of Incompressible Maxwell Fluid}

In terms of the velocity vector, the Cauchy balance can be rewritten as follows:
\begin{equation}
\mu \DD{\bm{v}}{t} + \mu \bm{v}\cdot \nabla \bm{v} = \diverg \hat \sigma = -\nabla p - \bm{E},
\label{eq:Cauchy1}
\end{equation}
where the left-hand side is the material (convective) derivative of the velocity vector $\bm{v}$  in the current configuration (called `convective' or `total' derivative). Remember that in the referential description, it is just the partial time derivative. Note also that for an incompressible metacontinuum, the density is the same constant in both the referential and spatial descriptions and we denote it by $\mu$. 
The concept of frame indifference (general covariance of the system) requires that the partial derivative of the stress variable (in our case the stress vector $-\bm{E}$) in Eq.~\eqref{eq:constitutive_solid_liquid} is replaced by the appropriate invariant rate. Since it is a vector density, see the argument by \citet{Schrodinger}, the rate has to leave the integral of the stress vector invariant. It is argued in \citep{Chri_FP} that the pertinent invariant rate is the so-called Oldroyd upper-convected derivative, namely
\begin{equation}
\frac{\eth \bm{E}}{\eth t} \stackrel{\mathrm{def}}{=}\DD{\bm{E}}{t} + \bm{v}\cdot \nabla \bm{E}- \bm{E} \cdot \nabla \bm{v} + (\nabla\cdot\bm{v}) \bm{E} .
\label{eq:elfield_Oldroyd}
\end{equation}

Here we come to one of the most crucial assumptions of the present work, namely, the way the constitutive relation has to be written when a relaxation of the stress is present. It is usually assumed that the invariant rate to be added to the constitutive law of viscous liquids should be of the stress tensor, i.e. the upper-convected Oldroyd derivative (\cite{Oldroyd,Bird}):
\begin{equation}
\frac{\eth \sigma}{ \eth t} := \frac{\partial \sigma}{\partial t} + \bm{v}\cdot \nabla \sigma - \sigma \grad \bm{v} - (\grad \bm{v})^\intercal \sigma  + \sigma \diverg \bm{v}. \label{eq:Oldroyd_tensor}
\end{equation}
The problem with this constitutive conjecture is that the tensor $\sigma$ does not  play an independent role in the Cauchy balance equation. Rather, the deviatoric stress vector $-
\bm{E} := \diverg \sigma$ appears there. Then,  does Eq.~\eqref{eq:Oldroyd_tensor} ensure that the time rate of $\diverg \sigma$ is invariant? To find out, we take the operation div of  Eq.~\eqref{eq:Oldroyd_tensor}, arriving at
\begin{multline}
\diverg(\frac{\eth \sigma}{ \eth t} )= \frac{\partial \diverg \sigma}{\partial t} + \bm{v}\cdot \nabla (\diverg \sigma) \\ - (\grad \bm{v})^\intercal (\diverg \sigma) 
+(\diverg  \sigma) \diverg \bm{v} -(\nabla \nabla \bm{v} )\sigma\\
= \frac{\eth \diverg \sigma}{\eth t}-(\nabla \nabla \bm{v} )\sigma, \label{eq:Oldroyd_vector}
\end{multline}
which differs from the invariant time rate of $\diverg \sigma$ by the term $(-\nabla \nabla \bm{v}) \sigma$, the latter being the contraction of the third rank tensor of the repeated gradient (the Hessian) of $\bm{v}$ and the second rank stress tensor $\sigma$. In order to establish which  constitutive relation is correct (the one involving the stress tensor or stress vector), one has to have data for flows in which the Hessian of the velocity vector field is measured.  While for the case of elastic liquids it is still possible to devise such an experiment, electromagnetism clearly indicates that the constitutive relation at play is the one involving the stress vector. 

Here we propose an alternative formulation for models involving stress relaxation by replacing the partial time derivative of the stress vector with the invariant rate
\begin{equation}
% \mu\big[\DD{\bm{v}}{t} + \bm{v}\cdot \nabla \bm{v}\big]  = -\nabla \phi - \bm{E} 
%\label{eq:Cauchy1}\\
\DD{\bm{E}}{t}  + \bm{v}\cdot \nabla \bm{E} 
-\bm{E}\cdot \nabla \bm{v}  + \bm{E} \,(\nabla\cdot \bm{v})  +\kappa \bm{E}
 = \eta \nabla \times (\nabla \times \bm{v}),
 \label{eq:Max_Ampere1}
 \end{equation}
which was proposed in \citep{Chri_FP}, and then successfully applied to the generalization of the  Maxwell--Cattaneo model of second sound in \citep{Chri_MRC}.

Apart from providing insight into the possible constitution of the electromagnetic field, the above model has practical significance for the theory of Maxwell elastic fluids.  Note that we add the condition  of incompressibility (recall Eq.~\eqref{eq:solen} with the subscript ``0"  omitted):
 \begin{equation}
 \nabla\cdot \bm{v} = 0. \label{eq:incompress1}
\end{equation}

Collectively, we can term the system Eqs.~\eqref{eq:Cauchy1}, \eqref{eq:Max_Ampere1} and \eqref{eq:incompress1} as the `Frame Indifferent Incompressible Maxwell Fluid Model'. %It does not use  the notion of magnetic field, but it contains the pressure  $p$, which is one of the unknown functions. Similarly to the case of incompressible liquids, $p$ provides the flexibility needed to satisfy the continuity equation. 

%Just like vorticity function in the models of fluid motion, magnetic field $\bm{B}$ is an auxiliary function, introduced to avoid the implicit potential function  $p$.

\section{Eliminating the Stress Vector}

The alternative formulation of the constitutive relation based on the stress vector concept proposed here has a very important consequence: it allows the stress vector $-\bm{E}$ to be eliminated between Eqs.~\eqref{eq:Max_Ampere1} and \eqref{eq:Cauchy1} (when $\nabla\cdot \bm{v} = 0$), to obtain a system that does not contain the stress variable  :\vspace{-0.07in}
\begin{multline}
\mu \frac{\partial^2 \bm{v}}{\partial t^2} + 2 \mu \bm{v} \cdot \nabla \frac{\partial \bm{v}}{\partial t}
+ (\bm{v}\bm{v})\nabla\nabla \bm{v} 
 = -\frac{\eth \nabla p}{\eth t}  +
 \eta \nabla^2 \bm{v}, \label{eq:second_order}
\end{multline}
where the following notations are used
\begin{align}
\frac{\eth \nabla p}{\eth t} &: = - \frac{\partial \nabla p}{\partial t}  - \bm{v} \cdot \nabla (\nabla p)  - (\nabla p)\cdot \nabla \bm{v},
\\
(\bm{v}\bm{v})\nabla\nabla \bm{v} &:= \sum\nolimits_{i=1}^3 \sum\nolimits_{j=1}^3 v_i v_j \nabla_i \nabla_{\!j} \, \bm{v}.
\end{align}
Note that $\eth/\eth t$ is the upper convected derivative of a vector defined in  Eq.~\eqref{eq:elfield_Oldroyd}.

Eq.~\eqref{eq:second_order} contains the implicit function $p$ that has to ensure the satisfaction of the incompressibility condition Eq.~\eqref{eq:incompress1}.  The advantage here is that, the only unknown functions are $\bm{v}$ and $p$, and the hyperbolicity of the model is now easily seen.

\section{Frame Indifferent Electromagnetodynamics}

According to the principal of frame indifference \citep{Truesdell}, the laws of physics (including the laws of continuum physics) must have the same form in any reference frame (coordinate system).  Unlike what is called `Lorentz Covariance', the laws in the  referential description are \emph{frame indifferent}, i.e. they are truly covariant.  However, experimental measurements are \emph{always} connected with a current frame in the geometric space. This means that an observational frame cannot detect the material variables (the referential description), but rather can merely measure their counterparts in the current (geometric) frame. This is a typical situation in mechanics of continuous media where the reference configuration is often not related to any measurable frame. For this reason we need to reformulate the model from Section~\ref{sec:Maxwell} in the current description making use of Euler variables. This is the objective of the present section.

 The Cauchy balance, Eq.~\eqref{eq:Cauchy1}, can be rewritten in the so-called `Gromeka--Lamb form' \citep{Sedov}:
\begin{equation}
\mu \DD{\bm{v}}{t} - \mu\bm{v}\times (\nabla \times \> \bm{v} )= -\nabla\phi  - \bm{E}, \quad \phi := p + \tfrac{1}{2}\bm{v}^2,  \label{eq:Lamb}
\end{equation}

Now, taking the \emph{curl} of Eq.~\eqref{eq:Lamb}, and using our definitions Eq.~\eqref{eq:elmag_definitions}, we get:
\begin{equation}
\nabla \times [ \bm{E}- \bm{v} \times  \bm{B}] = - \DD{\bm{B}}{t} ,
\label{eq:Faraday_Lorentz}
\end{equation}
which is Faraday's law with an additional term representing the force exerted by a moving magnetic field on each point of the medium. It is induced by the convective part of the acceleration at that point. The reaction to this force is the force acting on a moving point (charge), $\bm{v}\times \bm{B}$, known as the `Lorentz force'. In other words, the material invariant version of Faraday's law presented here automatically accounts for the physical mechanism that causes the Lorentz force. The latter is nothing more than the inertial force given by the convective part of the total derivative. This is a very important result because it tells us that the Lorentz force is not an additional, empirically observed force that has to be grafted onto Maxwell's equations, but rather, it is connected to the material time derivative, specifical, to its convective part.  Under the incompressibility condition,  Eq.~\eqref{eq:Faraday_Lorentz} can be recast as
\begin{equation}
\DD{\bm{B}}{t} + \bm{v}\cdot \nabla \bm{B} -\bm{B}\cdot\nabla \bm{v} = - \nabla \times\>  \bm{E},
\label{eq:Faraday_Hertz}
\end{equation}
which we can call the `Hertz form' of the Faraday--Lorentz law. Note the presence of the third term on the left-hand-side. It is not in Hertz's formulation, nor does it appear in \citep{Chri_FP,Chri_ML}. Evidently, Eq.~\eqref{eq:Faraday_Hertz} does not give any special advantage over Eq.~\eqref{eq:Faraday_Lorentz}, but the form of Eq.~\eqref{eq:Faraday_Hertz} shows that one cannot just add the convective part of the derivative to Faraday's law to make it frame indifferent. This follows from the fact that the magnetic field is not a primary variable (primary variables in fluids are the velocity components and the pressure). Rather, it is proportional to the $curl$ of the velocity vector, $\bm{v}$. 

By using the vector identity (see, e.g., \cite[pg.~180]{BoriTarap}
\begin{equation} \nabla \times (\bm{v} \times \bm{E}) = \bm{E}\cdot \nabla \bm{v} - \bm{v}\cdot \nabla \bm{E} + \bm{v} (\nabla\cdot \bm{E}) - \bm{E} (\nabla \cdot \bm{v}),
\label{eq:identity}
\end{equation}
%and acknowledging that $\nabla\cdot \bm{v}=0$, 
Eq.~\eqref{eq:Max_Ampere1} yields the  following generalization of the second of the dynamical  Maxwell's equation: %\begin{subequations}\label{eq:FIMIMF}
\begin{equation}
\DD{\bm{E}}{t}  - \underbrace{\nabla \times (\bm{v} \times \bm{E})} + \underline{\kappa \bm{E}} = - \widehat{\bm{v} (\nabla \cdot \bm{E})} + \widehat{\underbrace{ \underline{c^2 \nabla \times  \bm{B}}}},
 \label{eq:Max_Ampere}
\end{equation}
where the definition given in Eq.~\eqref{eq:H-def}  is already acknowledged.

Now, Eqs.~\eqref{eq:Faraday_Lorentz}, \eqref{eq:Max_Ampere}, \eqref{eq:H-def}, \eqref{eq:B_diverge}  form a system which can be termed the \emph{Frame Indifferent Electromagnetodynamics} (FIEM).  This system generalizes Hertz's program from 1890 and \emph{rigorously} fulfills the requirements for `General Covariance,' because FIEM is frame-indifferent; it is invariant when changing to another coordinate frame that can accelerate and can even deform. A very simple, limiting case of frame indifference is that of Galilean invariance.  The vector of absolute velocity $\bm{v}$ is the primary variable, but the absolute velocity in the referential description cannot be measured (only relative velocities can be observed). In principle it can be restored from the magnetic field Eq.~\eqref{eq:H-def}, provided a boundary condition is known. 

Clearly,  in the limit of small velocities the convective and convected terms can be neglected and in the limit one obtains Maxwell's system.

%It is important to impose the last equation \eqref{eq:incompress1} because it was used in the derivation of Eq.~\eqref{eq:Faraday_Lorentz}.

\section{Interpretation of the Different Terms in the Convected Time Rate}

A remarkable feature of Eq.~\eqref{eq:Max_Ampere} is that it incorporates terms (with under-braces) that, collectively, form the progenitor of  the Biot--Savart law. Indeed, setting aside the possible singularities connected  to point charges, then we can neglect the term with $\nabla\cdot \bm{E}$. Let us also consider the case of stationary electric field, $\bm{E}_t=0$, and no resistance $\kappa =0$. Then, considering a surface $D$ inside a closed contour $C$, we can integrate Eq.~\eqref{eq:Max_Ampere} over $C$ and use the Stokes theorem to obtain
\begin{equation}
\iint_D  [\bm{v} \times  \bm{E} + \bm{B}]\dd s =0, 
\end{equation}
where $\dd s$ is an areal element on the surface $D$. Since the surface $D$ is arbitrary, then the only possibility is that the integrand must be equal to zero. Thus one arrives at the form of the Biot--Savart law as it is stipulated in relativistic electrodynamics \cite[\S12.3]{Griffiths}, 
\begin{equation}
\bm{B} = -\frac{1}{c^2}\bm{v} \times \bm{E}
\end{equation}
where $\bm{v}$ is the velocity of the point of the field at which $\bm{B}$ is measured.

The underlined terms in Eq.~\eqref{eq:Max_Ampere} give that in \emph{vacuo}
\begin{equation}
\nabla \times \bm{B} = \frac{\kappa}{c^2} \bm{E},
\end{equation}
which is discussed in \cite[Ch.15]{Joos} in dimensional form. This can be interpreted as a combination of Ohm's and Ampere's laws of electromagnetism in \emph{vacuo}.

The terms with hats over them in Eq.~\eqref{eq:Max_Ampere} give as a corollary the Ampere law  if the following definition of a charge in \emph{vacuo} is introduced:
\begin{equation}
\rho \stackrel{\mathrm{def}}{=} \nabla\cdot \bm{E}. \label{eq:charge_def}
\end{equation}
In particular, the `chargedness', $\rho$, of the displacement/velocity field is defined as the divergence of the electric field at the point. In order not to confuse this property of the field \emph{in vacuo} with the localized pattern, called electron or a proton, we call the above defined function $\rho$ the \emph{metacharge}.  For the latter, a continuity equation is readily derived upon applying the operation $\diverg$ to Eq.~\eqref{eq:Max_Ampere}, namely
 \begin{equation}
 \DD{\rho}{t} + \nabla\cdot(\rho\bm{v} ) + \kappa \rho =0, 
 \label{eq:charge_continu}
  \end{equation}
which, for the case $\kappa=0$, is the standard continuity equation for charge. Here, is to be mentioned that Eq.~\eqref{eq:charge_continu} was derived by \citet{Chri_FP} directly from the Oldroyd form, Eq.~\eqref{eq:elfield_Oldroyd}, but the derivation here is much more straightforward because of the application of the identity Eq.~\eqref{eq:identity}. We shall refer to Eq.~\eqref{eq:charge_continu} as the continuity equation for the \emph{metacharge}.

In terms of the above introduced \emph{metacharge}\footnote{Note that the ubiquitous notion of charge can be explained away by a localized shear pattern in the metacontinuum, and then the usual charge is the integral of the metacharge over the span of the localized pattern (see \citet{Chri_ML}).}, we  obtain
\begin{equation}
c^2 \nabla \times \bm{B} = \rho \bm{v} = \bm{J},
\end{equation}
which can be called Ampere's law in \emph{vacuo}.

The important conclusion form the frame-indifferent formulation of the displacement current is that similarly to the Lorentz--force law, the convective/convected terms are related to phenomena that are  embodied in Ampere's and Biot--Savart's laws, thus unifying them with Maxwell's electrodynamics. All three electromagnetic-force laws (called alternatively the `laws of motional electromagnetism') are manifestations of the inertial forces in the metacontinuum.

 \citet[pp.~85--88]{Whittaker1} put forward the idea that both of these laws may actually follow from a single law, similar to what is presented in Eq.~\eqref{eq:Max_Ampere}. A  debate is still ongoing in the literature about wether these two laws are identical (see, e.g., \citep{Christodoulides,Jolly}) or independent (see, e.g., \cite{Graneau}). Our results seem to favor Whittaker's  original idea that both laws have to be interwowen in the correct formulation. In our work they are merely the corollaries form the inertial terms embodied in the convected derivative.

The frame indifferent model of the electromagnetic field (called here the metacontinuum), succeeds in unifying, in a single nexus, all known phenomena of electromagnetism: Faraday's law, displacement-current law, Lorentz-force law, Oersted--Ampere's law, and Biot--Savart law. It is a significant step forward from Maxwell's model, in which only the first two were explained by the equations themselves, and the latter three appeared as additional empirically observed relations between the main characteristics of the field. Since in our model these new terms are valid \emph{in vacuo}, they are clearly the progenitors of the related phenomena in moving media.

\section{Effect of Compressibility (Dark Energy?)}

As shown above, the model that yields Maxwell's equations is one of an incompressible elastic fluid. The obvious way to extend the validity of the model is to assume that the fluid is compressible. This means that it has to be a medium with vanishing compressibility, which happens when the dilational elasticity or viscosity coefficient is extremely large in comparison with the shear viscosity coefficient. As already mentioned, at this stage it is not clear whether the medium behaves as a liquid or solid when the compressional/dilational motions are considered. This question cannot be answered without staging an experiment in which the fabric of the metacontinuum could eventually be `ripped' in order to settle this question. Actually, if the  compressional/dilational motion is oscillatory, then it does not really make much difference if the metacontinuum is a solid or a fluid. As shown in the precedence, it has to be a fluid under shearing, but this does not impose any restriction on its behavior in the oscillatory  compression/dilation motions. Now, combining the two parts of the constitutive relation as given by Eqs.~\eqref{eq:dilational} the momentum equation can be written as
\begin{equation}
 \mu\big(\DD{\bm{v}}{t} + \bm{v}\cdot \nabla \bm{v}\big)  =- \bm{E} +   \begin{cases}
(\nu + 2 \zeta) \diverg \bm{v} & \text{liquid}, \\
(\lambda + 2 \eta) \diverg \bm{u} & \text{solid},
\end{cases} 
\label{eq:Cauchy2}
\end{equation}
which is to replace Eq.~\eqref{eq:Cauchy1}. Eq.~\eqref{eq:Max_Ampere1} remains unchanged, while the incompressibility condition, Eq.~\eqref{eq:incompress1}, has to be replaced by the following:
\begin{equation}
\frac{\partial{\mu}}{\partial t} + \bm{v} \cdot \nabla \mu = - \mu \diverg \bm{v}.  
\label{eq:compress}
\end{equation}
Alternatively, one can use the continuity equation in the from $\mu = J^{-1}\mu_r$ (see \cite{Chadwick,Truesdell}), where $J$ is the determinant of the gradient of deformation tensor  and $\mu_r$  is the \emph{constant} density in the referential description. Note that, the form chosen above, is consistent with the Eulerian description. Note also that if the solid rheology is assumed in Eq.~\eqref{eq:Cauchy2}, then one has to add the defining equation for the velocity components: $\bm{v} = \partial \bm{u}/ \partial t$,

Eqs.~\eqref{eq:Max_Ampere1}, \eqref{eq:Cauchy2},  and \eqref{eq:compress} form a coupled system for the compressible metacontinuum, which is highly nonlinear (even for the linear rheology) because of the dependence of the density $\mu$ on the motion. There are no conceptual difficulties to limit the model to the case of slight compressibility of the metacontinuum, and look for the propagation of harmonic compression waves therein. This raises the question about speed of the compressional waves (`sound') of the metacontinuum. In order to avoid ambiguous terminology we will not use the term \emph{sound} when referring to these  waves. Rather we will borrow a coinage from the ancient school of Stoa (see \cite{Sambursky}) and will call the compressional/dilational motions the \emph{pneuma}. 

One obvious implication of the existence of waves of a different kind than the electromagnetic (shear) waves is that, in fact, there is more energy in the physical vacuum than that  detected from electromagnetic interactions. Mechanically speaking, the \emph{pneuma} waves are `orthogonal' to the shear (EM) waves and they may not be detectable by devices based on  electromagnetic interactions.  Indeed, they perfectly fit the bill of what is currently called `dark energy' (see, e.g.,  \citet{HutererTurner}).  One can begin thinking of how to detect pneuma waves only after some solutions for coupled compressional and shear waves of the system Eqs.~\eqref{eq:Max_Ampere1}, \eqref{eq:Cauchy2},  and \eqref{eq:compress} become available. 

Let us mention in the conclusion of this section that the above derived system is suitable for any (visco)elastic medium, such as the ager-gelatin base phantom (see \cite{Catheline}).

\setcounter{equation}{0}
\section{Conclusion: The Rational Continuum Mechanics of Electromagnetic Field}

The approach proposed here presents a self-consistent point of view based on the continuum mechanics of the electromagnetic field whose shear deformations are perceived as the  phenomena of electromagnetism. It is shown that the linearized governing equations of any incompressible elastic medium admit a `Maxwell form'.  Conversely, the Maxwell equations of electrodynamics are a \emph{strict} mathematical corollary of the linearized governing equations of the incompressible Maxwell (visco)elastic fluid. 

This idea is further elaborated upon by deriving the frame indifferent formulation of the model of elastic fluids. It is argued that in some cases (as in the formulation of electromagnetism considered here and many other technologically significant applications) the constitutive relation can be written for the stress vector rather than for the stress tensor, because what actually enters the momentum equations is the point-wise stress vector, which is the divergence of the stress tensor. 

From the frame indifferent governing system of elastic fluids, a `Maxwell form' is derived that includes the terms of the original Maxwell equations along with terms stemming from the convective and convected invariant time rates. These inertial terms are the progenitors of the so-called laws of `motional electrodynamics': Biot--Savart, Oersted--Ampere, and Lorentz--force laws. The latter are usually assumed as additional empirical hypotheses to Maxwell's equations. This makes for a unified model of electromagnetism, which is truly covariant, by virtue of the fact that they are frame indifferent. In other words, the new model is invariant to changes to frames that accelerate and deform.

The continuum-mechanics formulation of the electromagnetic field proposed here opens a new avenue of research connected with the possible compressibility of  elastic fluids. Consequently two kinds of linear propagating waves can co-exist:  shear waves (light, when in the visible spectrum) and compressional waves (called \emph{pneuma} in this paper). It must be stressed that the speed of compressional waves is necessarily much larger than the speed of light. As a consequence, the metacontinuum appears virtually incompressible to observes using tools designed to detect its shear motions (i.e, electromagnetic phenomena).

\bigskip
\noindent{\bf References}
%\newskip\bibparskip
%\bibparskip -20mm
%\bibliography{NSF_analysis}

\end{document}